\documentclass{llncs}

\usepackage{color}

\usepackage{fullpage}

\usepackage{latexsym}
\usepackage{color,graphics}
\usepackage{color}
\usepackage{graphicx}
\usepackage{pgf}
\usepackage{tikz}
\usepackage{amsmath}
\usepackage{epsfig}

\usepackage{amssymb}
\usepackage{xspace}
\usepackage{ifthen}
\usepackage[T1]{fontenc}
\usepackage[latin9]{inputenc}
\usepackage{listings}
\usepackage{parcolumns}
\usepackage[english]{babel}

\newcommand{\junk}[1]{}

\newcommand{\cmt}[1]{\ifhmode\newline\fi{\sf *** \ \ #1 \\}}

\newtheorem{observation}{Observation}

\usepackage{multicol}

\renewcommand{\textcolor}[2]{#2}

\begin{document}

\title{\textcolor{red}{Node Sampling using Random Centrifugal Walks}
%\thanks{This research was supported in part by Comunidad de Madrid grant S2009TIC-1692 and Spanish MICINN grant TIN2008--06735-C02-01.}
}

%\author{\IEEEauthorblockN{Antonio Fern\'andez Anta}
%\IEEEauthorblockA{Institute IMDEA Networks, Madrid, Spain\\
%Email: antonio.fernandez@imdea.org}
%\and
%\IEEEauthorblockN{Andr\'es Sevilla}
%\IEEEauthorblockA{Dpto. Inform\'atica Aplicada \\ U.~Polit\'ecnica de Madrid, Madrid, Spain\\
%Email: asevilla@eui.upm.es}
%\and
%\IEEEauthorblockN{Alberto Mozo}
%\IEEEauthorblockA{Dpto. Arquitectura y Tecnolog\'ia de Computadores \\U.~Polit\'ecnica de Madrid, Madrid, Spain\\
%Email: amozo@eui.upm.es}
%}

%\author{\IEEEauthorblockN{Antonio Fern\'andez Anta\IEEEauthorrefmark{1},
%Andr\'es Sevilla\IEEEauthorrefmark{2},
%Alberto Mozo\IEEEauthorrefmark{3}} \\
%\IEEEauthorblockA{\IEEEauthorrefmark{1}Institute IMDEA Networks,
%Madrid, Spain\\
%Email: antonio.fernandez@imdea.org} \\
%\IEEEauthorblockA{\IEEEauthorrefmark{2}Dpto. Inform\'atica Aplicada,\\
%U.~Polit\'ecnica de Madrid, Madrid, Spain\\
%Email: asevilla@eui.upm.es} \\
%\IEEEauthorblockA{\IEEEauthorrefmark{3}Dpto. Arquitectura y Tecnolog\'ia de Computadores,\\
%U.~Polit\'ecnica de Madrid, Madrid, Spain\\
%Email: amozo@eui.upm.es}
%}

\author{
Andr\'es {\sc Sevilla}\inst{1}
	\and
	Alberto  {\sc Mozo}\inst{2}
	\and	
	Antonio {\sc Fern\'andez Anta}\inst{3}
}

\institute{
Dpto Inform\'atica Aplicada, U.~Polit\'ecnica de Madrid, Madrid, Spain\\
\email{asevilla@eui.upm.es}
\and
Dpto Arquitectura y Tecnolog\'ia de Computadores, U.~Polit\'ecnica de Madrid, Madrid, Spain\\
\email{amozo@eui.upm.es}
\and
Institute IMDEA Networks, Madrid, Spain\\
\email{antonio.fernandez@imdea.org}
%\and
%DIATEL, U.~Polit\'ecnica de Madrid, Madrid, Spain\\
%\email{jllopez@diatel.upm.es}
%\and
%\and
%LADyR, GSyC, U.~Rey Juan Carlos, M\'ostoles, Spain\\
%\email{anto@gsyc.es}
}

\date{}

\maketitle

\begin{abstract}

Sampling a network with a given probability distribution has been identified as a useful operation.
% to build network overlays. 
%For example, sampling nodes with uniform probability is the cornerstone of epidemic information spreading, while 
%constructing small-world network topologies and landmark-less positioning systems can be done by sampling with 
%a probability distribution that depends on the distance to a given node.
%
In this paper we propose distributed algorithms for sampling networks, 
so that nodes are selected by a special node, called the \emph{source}, 
with a given probability distribution. All these algorithms are based on a new class of random walks, that we 
call \emph{\textcolor{red}{Random Centrifugal Walks}} (RCW). A RCW is a random walk that starts at the source 
and \emph{always} moves \emph{away} from it. 

Firstly, an algorithm to sample any connected network using RCW is proposed. The algorithm assumes that each node has a weight, 
so that the sampling process must select a node with a probability proportional to its weight. This algorithm requires a 
preprocessing phase before the sampling of nodes.
In particular, a minimum diameter spanning tree (MDST) is created in the network, and then nodes' weights are efficiently aggregated using the tree. 
%(There are algorithms available to create the MDST, and weight aggregation can be done efficiently.) 
The good news are that the preprocessing is done only once,
\textcolor{red}{regardless} of the number of sources and the number of samples taken from the network.
After that, every sample is done with a RCW whose length is bounded by the network diameter.

Secondly, RCW algorithms that do not require preprocessing are proposed for grids and networks 
with regular concentric connectivity, for the case when the probability of selecting a node is a 
function of its distance to the source.

%We propose a RCW algorithm for connected networks
%%, because this family of networks includes most of the potential interesting networks. RCW 
%that selects a node with any desired probability distribution.
%% (not necessarily distance based). 
%A drawback of this algorithm 
%%for these networks 
%is that it needs preprocessing.
%%a previous calculation process. 
%When the probability distribution is distance based (i.e., the probability of selecting a node is a function of its associated weight),
%variants of the RCW algorithm without preprocessing are proposed.

The key features of the RCW algorithms 
(unlike previous Markovian approaches) are that (1) they do not need to warm-up (stabilize),
(2) the sampling
always finishes in a number of hops bounded by the network diameter, 
and (3) it selects a node with the \emph{exact probability distribution}.

\end{abstract}

\section{Introduction}

Sampling a network with a given distribution has been identified as a useful operation in many contexts.
% to build network overlays. 
For instance, sampling nodes with uniform probability is the building block of epidemic information spreading \cite{DBLP:journals/jacm/KempeKD04,DBLP:journals/tocs/JelasityVGKS07}. Similarly, sampling with a probability that depends on the distance to a given node \cite{DBLP:conf/opodis/BonnetKR07,springerlink:10.1007/978-3-642-17653-1_3} is useful to construct small world network topologies \cite{citeulike:47,DBLP:conf/stoc/FraigniaudG10,DBLP:conf/edcc/BertierBKLPR10}.
Other applications that can benefit from distance-based node sampling are landmark-less network positioning systems like NetICE9 \cite{NetICE9}, which does sampling of nodes with special properties to assign synthetic coordinates to nodes. In a different context, currently there is an increasing interest in obtaining a representative (unbiased) sample from the users of online social networks
%from their social graphs 
\cite{DBLP:conf/infocom/GjokaKBM10}.
In this paper we propose a distributed algorithm for sampling networks with a desired probability distribution.

%In this paper we propose distributed algorithms for sampling networks with a given probability distribution, so that nodes are selected at a special node, called the \emph{source}.
%% (note that the uniform distribution is a special case of this distribution). 
%As underlying networks, we consider several families, which span most of the networks found in practice: (a) connected networks,  (b) grids with lattice distance, and (c) networks where nodes are positioned at integral distances from the source node.
%, and in which a node at distance $k$ has some neighbor at distance $k-1$, 
% in which each node knows its probability to be selected. 
%As underlying networks we assume either a grid topology, or an arbitrary topology in which nodes are positioned at discrete distances from a source node, or a spanning tree topology.
%???????QUITAR?????????? (We discuss how the algorithms can be used for other topologies.) 
%The basic technique used for sampling is a new class of random walks that we call \emph{Centrifugal Random Walks} (RCW).
%A RCW starts at the source and \emph{always} moves away from it. Then, a key feature of RCW algorithms is that they finish 
%in a bounded number of hops, since they cannot move forever. Moreover, they select a node with the exact desired probability distribution.

\vspace{-1em}
\subsubsection{Related Work}

One technique to implement distributed sampling is to use gossiping between the network nodes. Jelasity et al.~\cite{DBLP:journals/tocs/JelasityVGKS07} present a general framework to
implement a uniform sampling service using gossip-based epidemic algorithms. Bertier et al.~\cite{DBLP:conf/edcc/BertierBKLPR10} implement uniform sampling and DHT services using gossiping. As a side result, they sample nodes with a distribution that is close to Kleinberg's harmonic distribution (one instance of a distance-dependent distribution). Another gossip-based sampling service that gets close to Kleinberg's harmonic distribution has been proposed by Bonnet et al.~\cite{DBLP:conf/opodis/BonnetKR07}. However, when using gossip-based distributed sampling as a service, it has been shown by Busnel et al.~\cite{Busnel2011} that only partial independence \textcolor{red}{($\epsilon$-independence)} between \textcolor{red}{views (the subsets of nodes held at each node)} can be guaranteed without re-executing the gossip algorithm. \textcolor{red}{Gurevich and Keidar~\cite{Gurevich} give an algorithm that achieves $\epsilon$-independence in $O(n s \log n)$ rounds (where $n$ is the network size and $s$ is the view size)}.
%samples 

Another popular distributed technique to sample a network is the use of random walks \cite{Zhong:2006:RWB:1151374.1151386}. Most random-walk based sampling algorithms do uniform sampling \cite{DBLP:conf/hicss/AwanFJG06,DBLP:conf/infocom/GjokaKBM10}, usually having to deal with the irregularities of the network. Sampling with arbitrary probability distributions can be achieved with random walks by re-weighting the hop probabilities to correct the sampling bias caused by the non-uniform stationary distribution of the random walks. Lee et al.~\cite{Lee2012} propose two new algorithms based on Metropolis-Hastings (MH) random walks for sampling with any probability distribution. These algorithms provide an unbiased graph sampling with a small overhead, and a smaller asymptotic variance of the resulting unbiased estimators than generic MH random walks.

%, for instance using Metropolis-Hastings random walks \cite{Metropolis:1953,Hastings70}.

Sevilla et al.~\cite{springerlink:10.1007/978-3-642-17653-1_3} have shown how sampling with an arbitrary probability distribution can be done without communication if a uniform sampling service is available. In that work, as in all the previous approaches, the desired probability distribution is reached when the stationary distribution of a Markov process is reached. The number of iterations (or hops of a random walk)
required to reach this situation (the warm-up time) depends on the parameters of the network and the desired distribution, but it is not negligible. For instance, Zhong and Sheng \cite{Zhong:2006:RWB:1151374.1151386} found by simulation that, to achieve no more than 1\% error, in a torus of 4096 nodes at least 200 hops of a random walk are required for the uniform distribution, and 500 hops are required for a distribution proportional to the inverse of the distance.
Similarly, Gjoka et al.~\cite{Gjoka2011} show that a MHRW sampler  needs about 6K samples (or 1000-3000 iterations) to obtain the convergence to the uniform probability distribution.
In the light of these results, Markovian approaches seem to be inefficient to implement a sampling service, specially if multiple samples are desired.  % and requiring several diagnostic tools to detect the convergence to the uniform probability distribution. 

\vspace{-1em}
\subsubsection{Contributions}

In this paper we present efficient distributed algorithms to implement a sampling service.
The basic technique used for sampling is a new class of random walks that we call \emph{\textcolor{red}{Random Centrifugal Walks}} (RCW).
A RCW starts at a special node, called the \emph{source}, and \emph{always} moves \emph{away} from it.

All the algorithms proposed here are instances of a generic algorithm that uses the RCW as basic element. 
This generic RCW-based algorithm works essentially as follows. A RCW always starts at the source node. 
When the RCW reaches a node $x$ (the first node reached by a RCW is always the source $s$),
% at distance $d(x)$ of the source, 
the RCW stops at that node with a \emph{stay probability}. If the RCW stops at node $x$, 
then $x$ is the node selected by the sampling. If the RCW does not stop at $x$, it jumps to a neighbor of $x$.
To do so, the RCW chooses only among neighbors that are \textcolor{red}{farther} from the source than the node $x$. 
%that are at a larger distance from the source than $x$ . 
(The probability of jumping to each of these neighbors is not necessarily the same.)
In the rest of the paper we will call all the instances of this generic algorithm as \emph{RCW algorithms.}

%Initially, we consider the family of networks that includes \emph{all connected networks}.
Firstly, we propose a RCW algorithm that samples \emph{any} connected network with \emph{any}
probability distribution (given as weights assigned to the nodes).
%(not necessarily distance based). 
%The drawback of this approach is that,
Before starting the sampling, a preprocessing phase is required.
This preprocessing involves building
a minimum distance spanning tree (MDST) in the network\footnote{Using, for instance, the algorithm proposed by Bui et al.~\cite{Bui2004} \textcolor{red}{whose time complexity is $O(n)$.}}, and using this tree for efficiently aggregating the node's  weights.
%performing a flooding and a convergecast over the tree. 
As a result of the weight aggregation, each node has to maintain one numerical value per link, which will be used by the RCW later. 
%However, this is compensated by the facts that,
%the algorithm can be applied in virtually all networks, and that 
Once the preprocessing is completed, any node in the network can be the source of a sampling process, and
multiple independent samplings with the exact desired distribution can be efficiently performed.
Since the RCW used for sampling follow the MDST, they take at most $D$ hops (where $D$ is the network diameter). 
%The algorithm used for the tree construction is not important for the correctness of the RCW algorithm, but the
%diameter of the tree will be an upper bound on the length of the RCW (and hence possibly the sampling latency). 
%There are several well known distributed algorithms (see, e.g., \cite{Elkin2004} and the references therein) that can be used to build the spanning tree. 
%Hence, it is interesting to build a minimum diameter spanning tree (MDST) because, as mentioned, the length of the RCW is  upper bounded by the tree diameter.  One possible candidate
%to be used in our context is the one proposed by Bui et al.~\cite{Bui2004}.
% Additionally, if link failures are expected, the variation of the former algorithm proposed by Gfeller et al.~\cite{Gfeller2011} can be used.

%Also,  $D$  will have an important impact in the convergecast preprocessing, when RCW is used in connected networks that require a spanning tree to work. 

Secondly, when the probability distribution is distance-based and nodes are at integral distances (measured in hops) from the source,
%(i.e., the probability of selecting a node is a function of its distance to the source),
RCW algorithms without preprocessing (and only a small amount of state data at the nodes) are proposed.
In a \emph{distance-based probability distribution}
all the nodes at the same distance from the source node are selected with the same probability. 
(Observe that the uniform and Kleinberg's harmonic distributions are special cases of distance-based probability distributions.)
%We find that the grid topology used is very restrictive to be practical. We hence define a more general network family (that includes the grid).
%The networks in this family have all the nodes at integral distances, $1, 2, \ldots, R$, from the source node. Furthermore, 
In these networks, each node at distance $k>0$ from the source has neighbors (at least) at distance $k-1$. 
We can picture nodes at distance $k$ from the source as positioned on a ring at distance $k$ from the source. 
The center of all the rings is the source, and the radius of each ring is one unit larger than the previous one. 
%(so that the $k$th ring is at distance $k$ from the source). 
Using this graphical image, we refer the networks of this family as \emph{concentric rings networks}.
\footnote{Observe that \emph{every} connected network can be seen as a concentric rings network. For instance, by finding the breadth-first search (BFS) tree rooted at the source, and using the number
of hops in this tree to the source as distance.}

%%Some of the variants of the RCW algorithm we propose sample with a 
%%In these distributions, 
%%(Of course, it must hold that $\sum_k n_k \times p_k=1$, where $n_k$ is the number of nodes at distance $k$.) 
%
\junk{ %%% Removed by Antonio 2012-07-04
This topology can also be imposed in real networks. For instance, consider a radio network in which 
each node has a fixed position assigned (say, with a GPS).
Then, fixing a source node, the nodes in the $k$th concentric rings can be the nodes whose (Euclidean) 
distance to the source is in the interval $(k-1,k]$. 
If the communication radius is reasonably large,
%with respect to the distance unit, 
the requirements of the concentric rings topology model will be satisfied.
} %%% Removed by Antonio 2012-07-04

%When the network can be modeled as a concentric rings network we can avoid the above calculation processes (spanning tree building, flooding and convergecast over the tree). 

%We consider two subfamilies of concentric rings networks: (a) grids with lattice distance, and (b) a more realistic model where nodes are positioned at integral distances from the source node.

The first distance-oriented RCW algorithm we propose samples with a distance-based distribution in a network with grid topology.
In this network, the source node is at position $(0,0)$ and the lattice (Manhattan) distance is used.
This grid contains all the nodes that are at a distance no more than the radius $R$ from the source 
(the grid has hence a diamond shape\footnote{A RCW algorithm for a square grid is easy to derive from the one presented.}). The algorithm we derive
assigns a stay probability to each node, that only 
depends on its distance from the source. However, the hop probabilities depend on the
position $(i,j)$ of the node and the position of the neighbors to which the RCW can jump to. 
We formally prove that the desired distance-based sampling probability distribution is achieved.
Moreover, since every hop of the RCW in the grid moves one unit of distance away from the source, 
the sampling is completed after at most $R$ hops.

We have proposed a second distance-oriented RCW algorithm that samples with distance-based distributions in concentric rings networks 
 \emph{with uniform connectivity}. These are networks in which all the nodes in each ring $k$ have the 
same number of neighbors in ring $k-1$ and the same number in ring $k+1$. Like the grid algorithm, this
variant is also proved to finish with the desired distribution in at most $R$ hops, where $R$ is the number of rings.
%
%The algorithm for this subfamily is very simple and requires no 
%preprocessing. A simple extension has also been developed for networks with uniform connectivity at distance 2.
%The largest drawback of these algorithms is that they use the number of nodes in a ring to compute the stay probability.

Unfortunately, in general, concentric rings networks have no uniform connectivity. This case
is faced by creating, on top of the concentric rings network, an overlay network that
has uniform connectivity. In the resulting network, the algorithm for uniform connectivity can be used. 
We propose a distributed algorithm that, if it completes successfully, builds the
desired overlay network. We have found via simulations that this algorithm succeeds in building
the overlay network in a large number of cases.

In summary, RCW can be used to
implement an efficient sampling service because, unlike previous Markovian (e.g., classical random walks and epidemic) approaches, 
(1)~it always finishes in a number of hops bounded by the network diameter, 
(2) selects a node with the \emph{exact probability distribution}, and
(3) does not need warm-up (stabilization) to converge to the desired distribution.
Additionally, in the case that preprocessing is needed, this only has to be executed once,
independently on the number of sources and the number of samples taken from the network.

The rest of the paper is structured as follows. In Section~\ref{s:model} we introduce concepts and notation that will be used in the rest of the paper.
In Section~\ref{s:spt-rings} we present the RCW algorithm for a connected network.
In Sections~\ref{s:grid} and \ref{s:uniform} we describe the RCW algorithm on grids and concentric rings networks with uniform connectivity. 
In Section~\ref{s:sum-simulations} we present the simulation based study of the algorithm for concentric rings topologies without uniform connectivity. Finally, we conclude the paper in Section~\ref{s:conclusions}.

%In Section~\ref{s:model} we introduce concepts and notation that will be used in the rest of the paper.
%In Section~\ref{s:grid} we describe the RCW algorithm on a grid network, and in Section~\ref{s:uniform} we present  the RCW algorithm on a concentric rings topology with uniform connectivity.
%In the next section we present the RCW algorithm for the a connected network.
%Finally, in Section~\ref{s:sum-simulations} we present the simulation based study of the algorithm for concentric rings topology without uniform connectivity.
%show the summary of some experimental results using RCW to sample a uniform and a PID probability distributions.

\section{Definitions and Model}
\label{s:model}

%\vspace{-1em}
\subsubsection{Connected Networks}

In this paper we only consider connected networks. This family 
includes most of the potentially interesting networks we can find.
In every network, we use $N$ to denote the set of nodes and $n=|N|$ the size of that set. When convenient, 
we assume that there is a special node in the network, called the \emph{source} and denoted by $s$. 
We assume that each node $x \in N$ has an associated weight $w(x)>0$. Furthermore, each node \emph{knows} its own weight.
%
%We make the following assumptions about these networks.
%\begin{enumerate}
%\item
%Connectivity. There is a path between each pair of nodes.
%\item
%Weights. Each node $x \in N$ has an associated weight $w(x)>0$. Furthermore, each node \emph{knows} its own weight.
%\end{enumerate}
%
The weights are used to obtain the desired probability distribution $p$, so that the 
probability of selecting a node $x$ is proportional to $w(x)$.
Let us denote $\eta = \sum_{j \in N} w(j)$.  Then, the probability of selecting $x \in N$ is $p(x)=w(x)/\eta$.
(In the simplest case, weights are probabilities, i.e., $w(x)=p(x), \forall x$ and $\eta=1$.) 

\vspace{-1em}
\paragraph{RCW in Connected Networks}

As mentioned, in order to use RCW to sample connected networks, some preprocessing is done. This involves constructing a spanning tree in the network and
performing a weight aggregation process. 
After the preprocessing, RCW is used for sampling.
A RCW starts from the source.
%, jumping to one of its neighbors in the tree. 
When the RCW reaches a node $x \in N$, it selects $x$ as the sampled vertex with probability $q(x)$, which we call the \emph{stay probability}. 
If $x$ is not selected, a neighbor $y$ of $x$ in the tree is chosen, using for that a collection of \emph{hop probabilities} $h(x,y)$. The values of $q(x)$ and $h(x,y)$ are computed in the preprocessing and stored at $x$. The probability of reaching a node $x \in N$ in a RCW is called the \emph{visit probability}, denoted $v(x)$. 

%Then, at each hop of the RCW the next node is chosen from the neighbors in the tree that are one more hop away from the source.
%We define the stay probability $q(x)$ when reaching node $x$ and the visiting probability $v(x)$ of $x$.
%
%However, it is not necessarily satisfied that all nodes at the same number of hops in the tree from the source will have the same
%visiting probability nor the same stay probability. In fact, since the probability of a node is given by its weight, which does not depend on
%the spanning tree used, nodes at the same distance in the tree from the source will usually have different probabilities. 

\vspace{-1em}
\subsubsection{Concentric Rings Networks}

We also consider a subfamily of the connected networks, which we call \emph{concentric rings networks.} 
These are networks in which the nodes of $N$ are at integral distances from $s$. In these networks, no node is at a distance from $s$ larger than a radius $R$. 
For each $k \in [0,R]$, we use $\mathbb{R}_k \neq \emptyset$  to denote the set of nodes at distance $k$ from $s$, and $n_k=|\mathbb{R}_k|$. 
(Observe that $\mathbb{R}_0 = \{s\}$ and $n_0=1$.)
These networks can be seen as a collection of concentric rings at distances $1$ to $R$ from the source, which is the common center of all rings.
For that reason, we call the set $\mathbb{R}_k$ the \emph{ring at distance $k$}.
For each $x \in \mathbb{R}_k$ and $k \in [1,R]$, $\gamma_{k}(x)>0$ is the number of neighbors of node $x$ at distance $k-1$ from $s$ (which is only 1 if $k=1$),  and $\delta_{k}(x)$ is the number of neighbors of node $x$ at distance $k+1$ from $s$ (which is 0 if $k=R$). 

%We assume that, for all $k<R$, it holds $\delta_{k}(x)>0$, i.e., $x$ is connected to at least one node in ring $k+1$. \marginpar{borrar?}

%Observe that \emph{every} connected network can be seen as a concentric rings network. For instance, by finding the BFS tree rooted at the source, and using the number
%of hops in this tree to the source as distance.
The concentric rings networks considered must satisfy the additional property that the probability distribution is \emph{distance based}. This means that, for all $k \in [0,R]$,
every node $x \in \mathbb{R}_k$ has the same probability $p_k$ to be selected. We assume that each node $x \in \mathbb{R}_k$ knows its own $p_k$. These properties allow, in the subfamilies defined below, to avoid the preprocessing required for connected networks.

\vspace{-.5em}
\paragraph{Grids}

A first subfamily of concentric rings networks considered is the grid with lattice distances. In this network, the source is at position $(0,0)$ of the grid, and it contains all the nodes $(i,j)$ so that $i,j \in [-R,R]$ and $|i|+|j| \leq R$. For each $k \in [0,R]$, the set of nodes in ring $k$ is $\mathbb{R}_k = \{(i,j):|i|+|j| =k \}$. The neighbors of a node $(i,j)$ are the nodes $(i-1,j)$, $(i+1,j)$, $(i,j-1)$, and $(i,j+1)$ (that belong to the grid). 

\vspace{-.5em}
\paragraph{Uniform Connectivity}

The second subfamily considered is formed by the concentric rings networks with \emph{uniform connectivity}. 
These networks satisfy that
% the following property.
%In this topology the connectivity among the different nodes of the rings satisfies the following rules: 
\begin{equation}
\label{satisfy_cond}
\forall k \in [1,R], \forall  x, y \in \mathbb{R}_k,  \delta_{k}(x) = \delta_{k}(y)   \land \gamma_{k}(x) = \gamma_{k}(y).
\end{equation}
In other words, all nodes of ring $k$ have the same number of neighbors $\delta_{k}$ in ring $k+1$ and the same number of neighbors $\gamma_{k}$ in ring $k-1$. 
%(Recall that each node in ring $k$ is connected to at least one node in ring $k-1$ and at least one node in ring $k+1$, if $k<R$.)
%
%We will also consider a stronger version of uniform connectivity, in which the uniform connectivity holds at distances 1 and 2. Finally, we
%will consider general concentric rings networks in which there is no uniform connectivity.

\vspace{-.5em}
\paragraph{RCW in Concentric Rings Networks}

The behavior of a generic RCW was already described. 
In the algorithm that we will present in this paper for concentric rings networks we guarantee that, 
for each $k$, all the nodes in $\mathbb{R}_k$ have the same visit probability $v_k$ and the same stay probability $q_k$. 
A RCW starts from the source.
%, jumping to one of its neighbors (in the first ring). 
When it reaches a node $x \in \mathbb{R}_k$, 
it selects $x$ as the sampled vertex with stay probability $q_k$.
%, which we call the \emph{stay probability}. 
If $x$ is not selected, a neighbor $y \in \mathbb{R}_{k+1}$ of $x$ is chosen.
%, using for that a collection of \emph{hop probabilities} $h_{k}(x,y)$.
%The visit probability of a node $x \in \mathbb{R}_k$ is denoted $v_k$.
%The probability of reaching a node $x \in \mathbb{R}_k$ is called the \emph{visit probability}, denoted $v_k$. 
%(Observe that $q_k$ and $v_k$ are the same for all nodes in $\mathbb{R}_k$.)

%Also, we will only consider RCW that take hops of distance 1 or 2. 

The desired \emph{distance-based probability distribution} is given by the values $p_k$, $k \in [0,R]$, where it must hold that $\sum_{k=0}^R n_k \times p_k =1$. The problem to be solved is to define the stay and hop probabilities so that the probability of a node $x \in \mathbb{R}_k$ is $p_k$. 

\begin{observation}
\label{o-vqp}
If for all $k \in [0,R]$ the visit $v_k$ and stay $q_k$ probabilities are the same for all the nodes in $\mathbb{R}_k$, the RCW samples with the desired probability iff $p_k=v_k \cdot q_k$.
\end{observation}

\section{Sampling in a Connected Network}
\label{s:spt-rings}

%Sometimes, as formerly shown, the desired distribution probability is not achieved when using a concentric rings topology. 
In this section, we present a RCW algorithm that can be used to sample any connected network.
As mentioned, in addition to connectivity, it is required that each node knows its own weight. A node will be selected with probability proportional to its weight. 

% ELIMINADO EN LA REVISION PARA EURO PAR
%The proposed algorithm overcomes some drawbacks of the previously presented RCW algorithms.
%Firstly, it works for all connected networks, not only concentric rings networks. 
%Secondly, it does not require uniform connectivity. Thirdly, since the weights assigned to nodes
%do not necessarily depend on the distance, any arbitrary probability distribution
%can be implemented. 
% ELIMINADO EN LA REVISION PARA EURO PAR

%This also removes the need
%the above sampling algorithms have of knowing the number of nodes in the ring.

\vspace{-1em}
\subsubsection{Preprocessing for the RCW Algorithm} The RCW algorithm for connected networks requires some preprocessing which will be described now. This preprocessing has to be done only once for the whole network, independently of which nodes act as sources and how many samples are taken.

\vspace{-.5em}
\paragraph{Building a spanning tree.}
Initially, the algorithm builds a spanning tree of the network. A feature of the algorithm is that, if several nodes
want to act as sources for RCW, they can all share the same spanning tree. Hence only one tree for the whole network has to be built.
The algorithm used for the tree construction is not important for the correctness of the RCW algorithm, but the
diameter of the tree will be an upper bound on the length of the RCW (and hence possibly the sampling latency). There are several well known distributed algorithms (see, e.g., \cite{Elkin2004} and the references therein) that can be used to build the spanning tree. In particular, it is interesting to build a minimum diameter spanning tree (MDST) because, as mentioned, the length of the RCW is  upper bounded by the tree diameter. There are few algorithms in the literature to build a MDST. One possible candidate
to be used in our context is the one proposed by Bui et al.~\cite{Bui2004}. Additionally, if link failures are expected, the variation of the former algorithm proposed by Gfeller et al.~\cite{Gfeller2011} can be used.

\begin{figure}[t]
\begin{lstlisting}{}
task $\mathit{Weight\_Aggregation} (i)$
if $i$ is a leaf then $\label{LC_isLeaf}$
  send WEIGHT($w(i)$) to neighbor $x$ $\label{LC_isLeafF}$
  receive WEIGHT($p$) from neighbor $x$
  $T_i(x)$ := $p$
else
  repeat
    receive WEIGHT($p$) from $x \in neighbors(i)$
    $T_i(x)$ := $p$
    foreach $y \in neighbors(i) \setminus \{x\}$ do
      if received WEIGHT($\cdot$) from $neighbors(i) \setminus \{y\}$ then
        send WEIGHT($w(i) + \sum_{z \in neighbors(i) \setminus \{y\}} T_i(z)$) to $y$
    end foreach
  until received WEIGHT($\cdot$) from every $x \in neighbors(i)$
\end{lstlisting}
\vspace{-1em}
\caption{Weight aggregation algorithm. Code for node $i$.}
\label{f-weight-aggr}
\end{figure}

% OPODIS 2012---------------

\vspace{-.5em}
\paragraph{Weight aggregation.}
Once the spanning tree is in place, the nodes compute and store aggregated weights using the algorithm of 
Figure \ref{f-weight-aggr}. The algorithm executes at each node $i \in N$, and it computes in a distributed way
the aggregated weight of each subtree that can be reached following one of the links of $i$. In particular, for each
node $x$ that is in the set of neighbors of $i$ in the tree, $\mathit{neighbors}(i)$, the algorithm computes a value $T_i(x)$
and stores it at $i$. Let $(i,x)$ be a link of the spanning tree, then by removing the link $(i,x)$ from the spanning tree there are two subtrees. We denote by $stree(x,i)$ the subtree out of them that contains node $x$.

\begin{theorem}
After the completion of the Weight Aggregation algorithm (of Figure \ref{f-weight-aggr}), each node $i \in N$ will store, for each
node $x \in \mathit{neighbors}(i)$, in $T_i(x)$ the value $\sum_{y \in stree(x,i)} w(y)$.
\end{theorem}
\begin{proof}
Consider $stree(x,i)$ a tree rooted at $x$. We prove the claim by induction in the depth of this tree. The base case is when the tree has depth 1. In this case $x$ is a leaf and, from the algorithm, it sends to $i$ its weight $w(x)$, which is stored at $i$ as $T_i(x)$.
If the depth is $k>1$, by induction hypothesis $x$ ends up having in $T_x(y)$ the sum of the weight of the subtree $stree(x,y)$, for each $y \in \mathit{neighbors}(x) \setminus \{i\}$. These values plus $w(x)$ are added up and sent to $i$, which stores the resulting value as $T_i(x)$. 
\end{proof}

The values $T_i(x)$ computed in this preprocessing phase will later be used by the RCW algorithm to perform the sampling. We can bound now the complexity of this process in terms of messages exchanged and time to complete. We assume that all nodes start running the Weight Aggregation algorithm simultaneously, that the transmission of messages  takes one step, and that computation time is negligible.
%
%The proof of the following theorem can be found in the appendix.

\begin{theorem}
\label{t:weight-aggregation}
The Weight Aggregation algorithm (of Figure \ref{f-weight-aggr}) requires $2(n-1)$ messages to be exchanged, and completes after $D$ steps, where $D$ is the diameter of the tree.
\end{theorem}

\begin{proof}
It is easy to observe in the algorithm that one message is sent across each link in each direction. Since all spanning trees have $n-1$ links, the first part of the claim follows. 

The second claim can be shown as follows. Let us consider any node $i$ as the root of the spanning tree. Let $d$ be the largest distance in the tree of any node from $i$.
We show by induction on $k$ that all nodes at distance $d-k$ from $i$ have received the aggregated weight of their corresponding subtrees by step $k$.
The base case is $k=1$, which follows since the leaves at distance $d$ send their weights to their parents in the first step. Consider now any (non-leaf) node $j$ at distance $d-k+1$ from $i$. Assume that $y$ is the parent (at distance $d-k$) of $j$ in the tree rooted at $i$.
By induction hypothesis $j$ has received all the aggregated weights of the subtrees by step $k-1$. Then, when the latest such value was received from a neighbor $x$ (Line 8), the foreach loop (Lines 10-13) if executed. In this execution, the condition of the if statement at Line 11 is satisfied for $y$. Then, the aggregated weight $w(j) + \sum_{z \in neighbors(j) \setminus \{y\}} T_j(z)$ is sent to $y$ by step $k-1$. This value reaches $y$ in one step, by step $k$.
Then, $i$ receives all the aggregated weights by step $d$. Since the largest value of $d$ is $D$, the proof is complete.
\end{proof}

\begin{figure}[t]
\begin{lstlisting}{}[]
task $\mathit{RCW}(i)$
when RCW_MSG($s$) received from $x$ $\label{rx_drwmsg}$
  $\mathit{candidates}$ := $neighbors(i)\setminus \{x\}$
  with $probability$ $q(i)=\frac{w(i)}{w(i) + \sum_{z \in \mathit{candidates}} T_i(z)}$ do $\label{prob}$
    $select$ $node$ $i$ and $report$ $to$ $source$ $s$ $\label{prob2}$
  otherwise 
    $choose$ $a$ $node$ $y$ $\in \mathit{candidates}$ with $probability$ $h(i,y)=\frac{T_i(y)}{\sum_{z \in \mathit{candidates}} T_i(z)}$ $\label{prob3} \label{prob4}$
    send RCW_MSG($s$) $to$ $y$ $\label{sendtoy}$ $\label{RCWstartF}$
\end{lstlisting}
\vspace{-1em}
\caption{RCW algorithm for connected networks. Code for node $i$.}
\label{f-SPTWeightSource}
\end{figure}

% OPODIS 2012. with q(s)=0
%\begin{figure}[t]
%\begin{lstlisting}{}[]
%task $\mathit{RCW}($n: Node$)$
%when RCW_START received $\label{RCW_start}$ 
%  $choose$ $a$ $node$ $x$ $\in neighbors(n)$ $\label{RCW_choose}$ with $probability$ $\frac{T_n(x)}{\sum_{z \in neighbors(n)} T_n(z)}$ $\label{choose2}$$\label{RCW_chooseF}$
%  send RCW_MSG($n$) $to$ $x$ $\label{RCW_send}$
%when RCW_MSG($s$) received from $x$ $\label{rx_drwmsg}$
%  $\mathit{candidates}$ := $neighbors(n)\setminus \{x\}$
%  with $probability$ $\frac{w(n)}{w(n) + \sum_{z \in \mathit{candidates}} T_n(z)}$ do $\label{prob}$
%    $select$ $node$ $n$ and $report$ $to$ $source$ $s$ $\label{prob2}$
%  otherwise 
%    $choose$ $a$ $node$ $y$ $\in \mathit{candidates}$ $\label{prob3}$ with $probability$ $\frac{T_n(y)}{\sum_{z \in \mathit{candidates}} T_n(z)}$ $\label{prob4}$
%    send RCW_MSG($s$) $to$ $y$ $\label{sendtoy}$ $\label{RCWstartF}$
%\end{lstlisting}
%\caption{RCW algorithm. Code for node $n$.}
%\label{f-SPTWeightSource}
%\end{figure}

% OPODIS 2012---------------

\vspace{-1em}
\subsubsection{RCW Sampling Algorithm}

%Recall that a RCW starts at the \emph{source} and \emph{always} moves away from it. 
In this RCW algorithm (Figure~\ref{f-SPTWeightSource}) any node can be
the source. The spanning tree and the precomputed aggregated weights are used by any node to perform the samplings (as many as needed).
The sampling process in the RCW algorithm works as follows.
% A RCW always starts at the source node when it receives a $\mathit{RCW\_START}$ message (Line \ref{RCW_start}), then it chooses a node among their node neighbors with probability $T(x)/\eta$ (Lines \ref{RCW_choose}-\ref{RCW_chooseF}) and sends to it a $\mathit{RCW\_MSG}$ message (Line \ref{RCW_send}).
To start the process, the source $s$ sends a message $\mathit{RCW\_MSG}(s)$ to itself.
%A RCW always starts at the source node when it receives a  message (Line \ref{rx_drwmsg}) triggered by itself.
When the $\mathit{RCW\_MSG}(s)$ message is received by a node $i$ from a node $x$,
it computes a set of candidates for next hop in the RCW, which are all the neighbors of $i$ except $x$.
% at distance $d(x)$ of the source, 
Then, the RCW stops and selects that node with a \emph{stay probability} $q(i)=\frac{w(i)}{w(i) + \sum_{z \in \mathit{candidates}} T_i(z)}$ (Line \ref{prob}). 
%If the RCW stops at node $i$, then $i$ is the node selected by the sampling. 
If the RCW does not select $i$, it jumps to a neighbor of $i$ different from $x$.
To do so, the RCW chooses only among nodes $y$ in the set of candidates (that move away from $s$) using $h(i,y)=\frac{T_i(y)}{\sum_{z \in \mathit{candidates}} T_i(z)}$ as hop probability (Line \ref{prob3}). 

\vspace{-1em}
\subsubsection{Analysis}

We show now that the algorithm proposed performs sampling with the desired probability distribution.

\begin{theorem}
\label{drw-spt}
Each node $i \in N$ is selected by the RCW algorithm with probability $p(i)=\frac{w(i)}{\eta}$.
\end{theorem}
\begin{proof}
If a node $i$ receives the $\mathit{RCW\_MSG}(s)$ from $x$, let us define $\mathit{candidates}=\mathit{neighbors}(i)\setminus \{x\}$, and
$T(i)=w(i) + \sum_{z \in \mathit{candidates}} T_i(z)$.
We prove the following stronger claim: Each node $i \in N$ is visited by the RCW with probability $v(i)=\frac{T(i)}{\eta}$ and
selected by the RCW algorithm with probability $p(i)=\frac{w(i)}{\eta}$.
 
We prove this claim by induction on the number of hops from the source $s$ to node $i$ in the spanning tree.
The base case is when the node $i$ is the source $s$. In this case $x$ is also $s$, $\mathit{candidates}=\mathit{neighbors}(s)$, and $T(s)=\eta$.
Hence, $v(s)=\frac{T(s)}{\eta}=1$ and $q(s)=\frac{w(s)}{\eta}$, yielding $p(s)=\frac{w(s)}{\eta}$.

The induction hypothesis assumes the claim true for a node $x$ at distance $k$ from $s$,
and proves the claim for $i$ which is at distance $k+1$. We have that 
$
\mathrm{Pr} [\mathrm{visit}\ i] = v(x) \left(1-q(x)\right)\frac{T(i)}{T(x)-w(x)},
$
where $1-q(x)$ is the probability of not selecting node $x$ when visiting it, and $\frac{T(i)}{T(x)-w(x)}$ 
is the probability of choosing the node $i$ in the next hop of the RCW. 
The stay probability of $x$ and $i$ are $q(x)=w(x)/T(x)$ and $q(i)=w(i)/T(i)$, respectively (Line \ref{prob}). 
Then,
$
v(i)  =  \frac{T(x)}{\eta}\left(1-\frac{w(x)}{T(x)}\right)\frac{T(i)}{T(x)-w(x)} 
= \frac{T(x)}{\eta} \left(\frac {T(x) - w(x)}{T(x)}\right) \frac{T(i)}{T(x)-w(x)} 
= \frac{T(i)}{\eta}
%\begin{eqnarray*}
%v(y) & = & \frac{T(x)}{\eta}\left(1-\frac{w(x)}{T(x)}\right)\frac{T(y)}{T(x)-w(x)} 
%&=& \frac{T(x)}{\eta} \left(\frac {T(x) - w(x)}{T(x)}\right) \frac{T(y)}{T(x)-w(x)} 
%&=& \frac{T(y)}{\eta}.
%\end{eqnarray*}
$
and
$
\mathrm{Pr} [\mathrm{select}\ i] = v(i) q(i) = \frac{T(i)}{\eta}\frac{w(i)}{T(i)}=\frac{w(i)}{\eta}.
$
\end{proof}

\lstset{numbers=left,columns=fullflexible,tabsize=4,basicstyle=\small,identifierstyle=\rmfamily\textit,mathescape=true,morekeywords={repeat, forever,set,task,process,const,when,with,wait,
send, otherwise,elsif,procedure,in,null,function,do,loop,until,return,foreach,begin,var,if,then,else,rcv,rx_user_events,rx_network_events,for,each,end,received,upstream,downstream,while,do,forward,backward,to},literate={:=}{{ $\leftarrow$ }}1{->}{{ $\rightarrow$ }}1}

\lstset{escapeinside={('}{')}}

\section{Sampling in a Grid}
\label{s:grid}

\vspace{-0.5em}
If the algorithm for connected networks is applied to a grid, 
%When applying the described process to a grid, 
given its regular structure, the construction of the spanning tree could be done without any communication among nodes, but the weight aggregation process has to be done as before. However, we show in this section that all preprocessing
% (building the spanning tree, and weight aggregation) 
and the state data stored in each node 
can be avoided if the probability distribution is based on the distance. 
RCW sampling process was described in Section \ref{s:model}, and we only redefine stay and hop probabilities.  
%Only a small (and constant) amount of state data is stored in each node.
%
%In this section we present a RCW algorithm variant for the grid. 
%
From Observation~\ref{o-vqp}, the key for correctness is to assign stay and hop probabilities that guarantee visit and stay probabilities that are homogenous for all the nodes at the same distance from the source.

\vspace{-1.5em}
\subsubsection{Stay probability} For $k \in [0,R]$, the stay probability of every node $(i,j) \in \mathbb{R}_k$ is defined as
%
%\[
%  q_k = \left\{ 
%  \begin{array}{l l}
%    n_k \cdot p_k & \quad \text{if $k=0$}\\
%    \\
%    \frac {n_k \cdot p_k}{\sum_{j=k}^{R} n_j \cdot p_j} & \quad \text{if $k > 0$}\\
%  \end{array} \right.
%\]
%where 
\begin{equation}
q_k = \frac {n_k \cdot p_k}{\sum_{j=k}^{R} n_j \cdot p_j} = \frac {n_k \cdot p_k} {1-\sum_{j=0}^{k-1} n_j \cdot p_j}.
\end{equation}

% when q(s)=0
%$$
%q_k = \frac {n_k \cdot p_k}{\sum_{j=k}^{R} n_j \cdot p_j} = \frac {n_k \cdot p_k} {1-\sum_{j=1}^{k-1} n_j \cdot p_j}.
%$$
As required by Observation~\ref{o-vqp}, all nodes in $\mathbb{R}_k$ have the same $q_k$. Note that $q_0=p_0$ and $q_R=1$, as one may expect. Since the
value of $p_k$ is known at $(i,j) \in \mathbb{R}_k$, $n_k$ can be readily computed\footnote{$n_0=1$, while $n_k=4 k$ for $k \in [1,R]$.}, and the value of $\sum_{j=0}^{k-1} n_j \cdot p_j$ can be piggybacked in the RCW, the
value of $q_k$ can be computed and used at $(i,j)$ without requiring precomputation nor state data.

%When a Drifting Random Walk arrives to a node, it can stop on it with a probability $q_i$ proportional to the distance $i$ travelled from the source node.
%We define this {\em stay probability} $q_i$ as:
%\\
%\\
%$q_i = \frac {P_y}{P_m+P_y}, \,\,where$
%\\
%\\
%$P_y = \sum_{j: d(i,j)=d(i,origin)} \,,p_j \, \,\,;  \,\,\, P_m = \sum_{j:d(i,j)>d(i,origin)} p_j
%\\
%\\
%q_i = \frac {n_i \cdot p_i}{n_i \cdot p_i + \sum_{j=i+1}^{max dist} n_j \cdot p_j} = \frac {n_i \cdot p_i} {1-\sum_{j=1}^{i-1} n_j \cdot p_j}$
%\\
%\\
%In each RCW step, $P_y$ is the sum of the $p_d$ of all the nodes that are at same distance $d$ from the source. Where $d$ is the distance from the source node to the node of the current RCW step. On the other hand, $P_m$ is the sum of the $p_j$ of all the nodes whose distance $j$ from source is greater than $d$.
%Note that the probability $q_i$ converges to $1$ as the distance from source to the current node of the RCW grows.

%\vspace{-1.0em}
\subsubsection{Hop probability}
In the grid, the hops of a RCW increase the distance from the source by one unit. We want to guarantee that the visiting probability is the same for each node at the same distance, to use
Observation~\ref{o-vqp}. To do so, we need to observe that nodes $(i,j)$ over the axes (i.e., with $i=0$ or $j=0$) have to be treated as a special case, because they can only be reached via a single path, while the others nodes can be reached via several paths. To simplify the presentation, and since the grid is symmetric, we give the hop probabilities for one quadrant only (the one in which nodes have both coordinates non-negative). The hop probabilities in the other three quadrants are similar. The first hop of each RCW chooses one of the four links of the source node with the same probability $1/4$.
We have three cases when calculating the hop probabilities from a node $(i,j)$ at distance $k$, $0<k<R$, to node $(i',j')$.
 
%To simplify the calculation we use a single quadrant among the four possible quadrants in a grid. We choose the quadrant where $x$ and $y$ coordinates are positive.

\begin{itemize}
\item \emph{Case A:} The edge from $(i,j)$ to $(i',j')$ is in one axis \textcolor{red}{(i.e., $i = i' = 0$ or $j = j' = 0$)}. The hop probability of this link is set to $h_{k}((i,j),(i',j')) = \frac {i+j}{i+j+1}= \frac {k}{k+1}$.

%The RCW follows the axes. The nodes at distance $k+1$ have only one neighbor at distance $k$. The probability to step from a node with coordinates $(i,j)$ at a distance $k$ to a node at distance $k+1$ is $h_{k+1} = \frac {i+j-1}{i+j}$.
%\\

\item \emph{Case B:} The edge from $(i,j)$ to $(i',j')$ is not in the axes, $i'=i+1$, and $j'=j$. The hop probability of this link is set to $h_{k}((i,j),(i+1,j)) = \frac {2i+1} {2(i+j+1)}=\frac {2i+1} {2(k+1)}$.
%The current RCW step is not in the axes. The current  node is at distance $k$ and RCW moves to the right  to coordinate $(i, j)$ at $k+1$ distance. The nodes at distance $k+1$ have two neighbors at distance $k$.\\

\item \emph{Case C:} The edge from $(i,j)$ to $(i',j')$ is not in the axes, $i'=i$, and $j'=j+1$. The hop probability of this link is set to $h_{k}((i,j),(i,j+1)) = \frac {2j+1} {2(i+j+1)}=\frac {2j+1} {2(k+1)}$.
%This case is the same as above, but in the Y axis moving up, so hop probability is $h_{k+1} = \frac {2j+1}{2(i+j+1)}$.
\end{itemize}
It is easy to check that the hop probabilities of a node add up to one.
%$d_M$ is the Manhattan distance between two nodes. $i$ and $j$ represent the grid coordinates.

%\vspace{-1em}
\subsubsection{Analysis}
In the following we prove that the RCW that uses the above stay and hop probabilities selects nodes with the desired sample probability.

\begin{lemma}
All nodes at the same distance $k \ge 0$ to the source have the same visit probability $v_k$.
\end{lemma}

%Note that: a) all nodes at distance $k$ have the same $q_k$ probability, b) RCW steps always increment the distance from source, c) we will use the former A, B and C cases.

\begin{proof}
%The proof uses induction. The base case is $k=1$, which is trivial since the probability of visiting each of the four nodes at distance 1 from the source $s$ is $v_i=1/4$.
The proof uses induction. The base case is $k=0$, and obviously $v_k = 1$. When $k=1$, the probability of visiting each of the four nodes at distance 1 from the source $s$ is $v_i=\frac{1-q_0}{4}$, where $1-q_0$ is the probability of not staying at source node.
Assuming that all nodes at distance $k >0$ have the same visit probability $v_k$, we prove the case of distance $k+1$. Recall that the stay probability is the same $q_k$ for all nodes at distance $k$.

The probability to visit a node $x=(i',j')$ at distance $k+1$ depends on whether $x$ is on an axis or not. If it is in one axis it can only be reached from its only neighbor $(i,j)$ at distance $k$. This happens with probability (case A)
$
\mathrm{Pr} [\mathrm{visit} \ x] = v_k (1-q_k) \frac {i+j}{i+j+1} = v_k (1-q_k) \frac {k}{k+1}.
$
If $x$ is not on an axis, it can be reached from two nodes, $(i'-1,j')$ and $(i',j'-1)$, at distance $k$ (Cases B and C). Hence, the probability of reaching $x$ is then
%\begin{multline}
$
\mathrm{Pr} [\mathrm{visit} \ x] = v_k (1-q_k) \frac {2(i'-1)+1} {2(i'+j')}  + v_k (1-q_k) \frac {2(j'-1)+1} {2(i'+j')} = v_k (1-q_k)\frac {k} {k+1}.
$
%\end{multline}
Hence, in both cases the visit probability of a node $x$ at distance $k+1$ is $v_{k+1}= v_k (1-q_k) \frac {k} {k+1}$. This proves the induction and the claim.
\end{proof}

%Observe that the algorithm never comes back, it moving always away from the source.

%\begin{claim}
%One node with distance equal or less than $k$ is chosen with probability $p_k$.
%\end{claim}

\begin{theorem} 
\label{teorema1}
Every node at distance $k \in [0,R]$ from the source is selected with probability $p_k$.
\end{theorem}

\begin{proof}
If a node is visited at distance $k$, it is because no node was selected at distance less than $k$, since a RCW always moves away from the source. 
Hence,
$
\mathrm{Pr}[\exists x \in \mathbb{R}_k\ \mathrm{visited}] =1-\sum_{j=0}^{k-1} n_j p_j.
$
Since all the $n_k$ nodes in $\mathbb{R}_k$ have the same probability to be visited (from the previous lemma),
we have that
$
v_k = \frac{1-\sum_{j=0}^{k-1} n_j p_j}{n_k}.
$
Now, since all the $n_k$ nodes in $\mathbb{R}_k$ have the same
stay probability is $q_k$, the probability of selecting a particular node $x$ at distance $k$ from the source is
%$q_k=\frac {n_{k} p_{k}} {\sum_{j=k}^{R} n_j p_j}$, 
%(given that node $x$ has the same probability to be visited than the other $n_{k+1} $ nodes).
%
%The probability of choosing a node $x$ with distance $k+1$ is equal to:
%
%\[ \mathrm{Pr} [choose \ x] = \frac {n_{k+1} p_{k+1}} {\sum_{j=k+1}^{\infty} n_j p_j}\]
%
%since the node $x$ has been visited with the same probability that the other nodes at a distance $k+1$
$
\mathrm{Pr} [\mathrm{select}\ x] = v_k q_k = \frac{1-\sum_{j=0}^{k-1} n_j p_j}{n_k} \frac {n_{k} p_{k}}{\sum_{j=k}^{R} n_j p_j} = p_k,
$
%\[
%  v_k = \left\{ 
%  \begin{array}{l l}
%    1 & \quad \text{if $k=0$}\\
%    \\
%    \left(1-\sum_{j=0}^{k-1} n_j p_j \right) & \quad \text{if $k > 0$}\\
%  \end{array} \right.
%\]
where it has been used that %$\sum_{j=0}^{R} n_j p_j = 1$ and that 
$(1-\sum_{j=0}^{k-1} n_j p_j) = \sum_{j=k}^{R} n_j p_j$.

% OPODIS 2012 when q(s)=0
%$$
%\mathrm{Pr} [\mathrm{select}\ x] = \left(1-\sum_{j=1}^{k-1} n_j p_j \right) \frac {1}{n_{k}} \frac {n_{k} p_{k}}{\sum_{j=k}^{R} n_j p_j} = p_k.
%$$
%Where it has been used that $\sum_{j=1}^{R} n_j p_j = 1$ and that $(1-\sum_{j=1}^{k-1} n_j p_j) = \sum_{j=k}^{R} n_j p_j$.
\end{proof}

\section{Sampling in a Concentric Rings Network with Uniform Connectivity}
\label{s:uniform}

In this section we derive a RCW algorithm to sample a concentric rings network with uniform connectivity, where all preprocessing is avoided, and only a small (and constant) amount of data is stored in each node.
Recall that uniform connectivity means that all nodes of ring $k$ have the same number of neighbors $\delta_{k}$ in ring $k+1$ and the same number of neighbors $\gamma_{k}$ in ring $k-1$.

\vspace{-1em}
\subsubsection{Distributed algorithm}

The general behavior of the RCW algorithm for these networks was described in Section \ref{s:model}.
In order to guarantee that the algorithm is fully distributed, and to reduce the amount of
data a node must know a priori, a node at distance $k$ that sends the RCW to a node in ring $k+1$
piggybacks some information. More in detail, when a node in ring $k$ receives the RCW from a node of ring $k-1$, 
it also receives the probability $v_{k-1}$ of the previous step, and the values $p_{k-1}$, $n_{k-1}$, and $\delta_{k-1}$. 
Then, it calculates the values of $n_k$, $v_k$, and $q_k$. 
After that, the RCW algorithm uses the stay probability $q_k$ to decide whether to select the node or not. 
If it decides not to select it, it chooses a neighbor in ring $k+1$ with uniform probability. 
Then, it sends to this node the probability $v_k$ and the values $p_k$, $n_k$, and $\delta_{k}$, 
%of the current ring 
piggybacked in the RCW. 

The RCW algorithm works as follows.
The source $s$ selects itself with probability $q_0=p_0$. If it does not do so, it chooses one node in ring $1$ with uniform probability,
and sends it the RCW message with values $v_0=1$, $n_0=1$, $p_0$, and $\delta_0$.
Figure \ref{f-drw} shows the code of the RCW algorithm for nodes in rings $\mathbb{R}_k$ for $k>0$.
Each node in ring $k$ must only know initially the values $\delta_k$, $\gamma_k$ and $p_k$.
Observe that $n_k$ (number of nodes in ring $k$) can be locally calculated as $n_k = n_{k-1} \delta_{k-1}/\gamma_{k}$. 
%Each node in ring $k$ receives $n_{k-1}$ and $\delta_{k-1}$ from the neighbors in ring $k-1$, and sends $n_k$ and $\delta_k$ to its neighbors in ring $k+1$. 
%The base case is $k=0$ and thus the source node knows $n_0 = 1$ and $\delta_0$. 
The correctness of this computation follows from the uniform connectivity assumption (Eq. \ref{satisfy_cond}).

\begin{figure}[t]
%\begin{minipage}[t]{1\columnwidth}
%
\begin{lstlisting}
task $\mathit{RCW} (x,k,\delta_k,\gamma_k,p_k)$
when $\mathit{RCW\_MSG}(s, v_{k-1}, p_{k-1}, n_{k-1},\delta_{k-1})$ received: $\label{alg-rec}$
   $n_k$ := $n_{k-1} \frac{\delta_{k-1}}{\gamma_{k}}$; $\ \ $  $v_k$ := $n_{k-1} \frac{v_{k-1} - p_{k-1}}{n_k}$; $\ \ $ $q_k$ := $\frac{p_k}{v_k}$ $\label{alg-comp}$
   with probability $q_k$ do select node $x$ and report to $s$
   otherwise
      choose a neighbor $y$ in ring $k+1$ with uniform probability
      send $\mathit{RCW\_MSG}(s, v_k, p_k, n_k, \delta_{k})$ to $y$
\end{lstlisting}
%
%\end{minipage}
\vspace{-1em}
\caption{RCW algorithm for concentric rings with uniform connectivity (code for node $x \in \mathbb{R}_k$, $k>0$).}
\label{f-drw}
\end{figure}

\vspace{-1em}
\subsubsection{Analysis} The uniform connectivity property can be used to prove by induction that all nodes in the same ring $k$ have the same probability $v_k$ to be reached.
The stay probability $q_k$ is defined as $q_k=p_k/v_k$. Then, from Observation~\ref{o-vqp}, the probability of selecting a node $x$ of ring $k$ is $p_k=v_k q_k$. What is left to
prove is that the value $v_k$ computed in Figure \ref{f-drw} is in fact the visit probability of a node in ring $k$.

\begin{lemma}
The values $v_k$ computed in Figure \ref{f-drw} are the correct visit probabilities.
\end{lemma}

\begin{proof}
Let us use induction. For $k=1$ the visit probability of a node $x$ in ring $\mathbb{R}_1$ is $\frac{1 - q_0}{n_1}=\frac{1 - p_0}{n_1}$.
On the other hand, when a message $\mathit{RCW\_MSG}$ reaches $x$, it carries $v_0=1$, $n_0=1$, $p_0$, and $\delta_0$ (Line~\ref{alg-rec}).
Then, $v_1$ is computed as $v_1= n_0 \frac{v_0 - p_0}{n_1} = \frac{1 - p_0}{n_1}$ (Line~\ref{alg-comp}).
%the visit probability sent by the source $s$ (the only node in ring $\mathbb{R}_0$) is $v_0=1$, which is the correct value of $v_0$.
 %piggybacked in a RCW message sent to a node in ring $1$.
% while the value computed by the algorithm is $v_1=n_0(v_0  - p_0)/n_1=1/n_1$.
For a general $k>1$, assume the value $v_{k-1}$ is the correct visit probability of a node in ring $k-1$. The visit probability of a node in ring $k$ is
$v_{k-1} n_{k-1} (1-q_{k-1})/n_k$, which replacing $q_{k-1}=p_{k-1}/v_{k-1}$ yields the expression used in Figure \ref{f-drw} to compute $v_k$ (Line~\ref{alg-comp}).
\end{proof}

%As previously defined, the probability to select a node $x$ at distance $i$ is:
%$$
%\mathrm{Pr} [select \ x] = p_i = Pr [visit \ ring \ i] \times \frac{1}{n_i} \times q_i
%$$
%Let define the probability to visit a ring $i$ in recurrent way:
%$$
%\mathrm{Pr} [visit \ ring \ i] = v_i =  Pr[visit \ ring \ i-1] \times  (1-q_{i-1})
%$$

%As random walk starts at ring $0$, we assign the probabilities $v_0=1$, and  $p_0=0$ for calculating the probability of the first step.
%Also, we define $B_i = (1-q_i)$ as the probability of not stop at ring $i$, then:
%$$
%v_i =  v_{i-1} \times B_{i-1} 
%$$
%If we substitute $B_{i-1}$ in the $v_i$ formula, we obtain the new expression of $v_i$ is
%%We define $B_i = \frac {v_i-n_i \times p_i}{v_i}$
%%The new expression of the probability of visit ring $i$ is
%$$
%v_i=(v_{i-1}-n_{i-1} \times p_{i-1}) 
%$$
%This expression only needs the information sent from a node at the previous step in the RCW.

The above lemma, together with the previous reasoning, proves 
the following.
% theorem.

\begin{theorem}
Every node at distance $k$ of the source is selected with probability $p_k$.
\end{theorem}

%A more general approach when nodes in a concentric rings network have uniform connectivity with neighbors at distance 1 and 2 can be found in Appendix \ref{s:largerdistances}.

%The proof of this theorem is the same as the former {\em Grid Topology} theorem \ref{teorema1}.

%Then, for calculating $v_i$ probability we only need to know the visit probability and the number of nodes of previous ring ($i-1$) 

%\begin{figure*}[t]
%%%\scalebox{0.4}{0.6}{\input{camino2.pstex_t}}
%%\begin{center}
%%\scalebox{0.16}{\includegraphics{imagenes/i_u_k1_k2}}
%\scalebox{0.14}[0.16]{\includegraphics{imagenes/sinAAP}}
%\scalebox{0.14}[0.16]{\includegraphics{imagenes/conAAP}}
%%\scalebox{0.13}{\includegraphics{imagenes/i_r_k1_k2}}
%\caption{UNI and PID scenarios without uniform connectivity. Without using the AAP algorithm (left side) and using it (right side).}
%%\vspace*{-5ex}
%\label{fig:E1-2}
%%\end{center}
%\end{figure*}
%
%\junk{
%\begin{figure*}[t]
%%%\scalebox{0.4}{0.6}{\input{camino2.pstex_t}}
%%\begin{center}
%\scalebox{0.13}{\includegraphics{imagenes/u_r_h_k1}}
%\scalebox{0.13}{\includegraphics{imagenes/i_r_h_k1}}
%\caption{Uniform distribution and PID scenarios in the Experiment 2 using the AAP algorithm.}
%%\vspace*{-5ex}
%\label{fig:E1-3}
%%\end{center}
%\end{figure*}
%}

\begin{figure}[t]
\begin{center}
\scalebox{0.5}[0.5]{\includegraphics{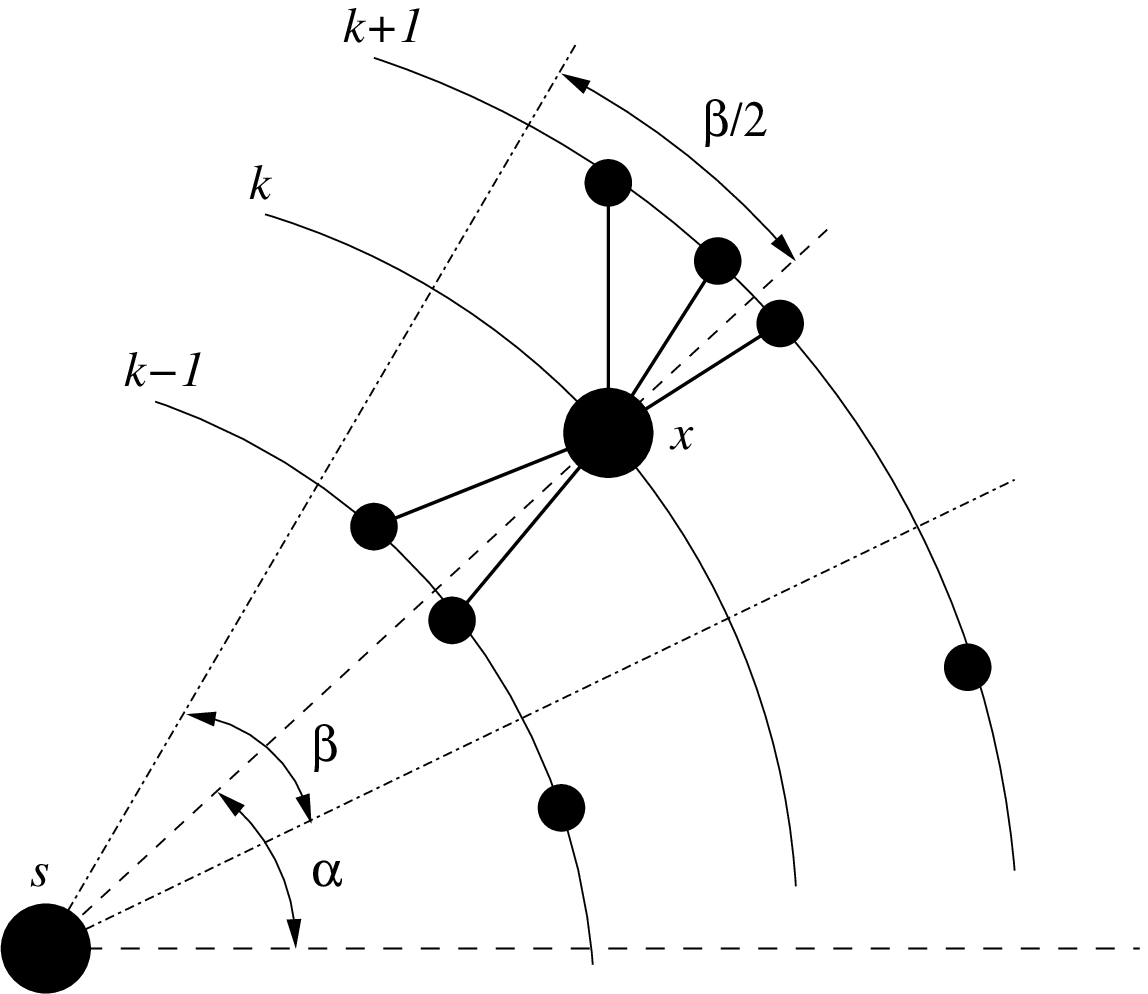}}
\end{center}
\vspace{-1.0em}
\caption{Node deployment and connectivity used in the simulations.}
\label{fig:simul}
\end{figure}

\section{Concentric Rings Networks without Uniform Connectivity}
\label{s:sum-simulations}

Finally, we are interested in evaluating, by means of simulations, the performance of the RCW algorithm for concentric rings with uniform connectivity when it is used on a more realistic topology: a concentric rings network \emph{without uniform connectivity.} 
The experiment has been done in a concentric rings topology of $100$ rings with $100$ nodes per ring, and it places the nodes of each ring uniformly {\em at random} on each ring.
This deployment does not guarantee uniform connectivity. \textcolor{red}{Instead, the nodes' degrees follow roughly a normal probability distribution.}
In order to establish the connectivity of nodes, we do a geometric deployment. A node $x$ in ring $k$ is assigned a position in the ring.
This position can be given by an angle $\alpha$. Then, each network studied will have associated a connectivity angle $\beta$, the same for all nodes.
This means that $x$ will be connected to all the nodes in rings $k-1$ and $k+1$ whose
position (angle) is in the interval $[\alpha-\beta/2, \alpha+\beta/2]$. (See Figure~\ref{fig:simul}.) Observe that the bigger the angle $\beta$ is, the more neighbors $x$ has in rings $k-1$ and $k+1$.
We compare the relative error of the RCW algorithm when sampling with two distributions: the uniform distribution (UNI) and a distribution proportional to the inverse of the distance (PID).  We define the relative error $e_i$ for a node $x$ in a collection $C$ of $s$ samples as $e_i=\frac{|fsim_x - f_x|}{f_x}$, where $fsim_x$ is the number of instances of $x$ in collection $C$ obtained by the simulator, and $f_x= p_x \cdot s$ is the expected number of instances of $x$ with the ideal probability distribution (UNI or PID). We compare the error of the RCW algorithm with the error
of a generator of pseudorandom numbers.
For each configuration, a collection of $10^7$ samples has been done.

\begin{figure*}[t]
\scalebox{0.22}[0.22]{\includegraphics{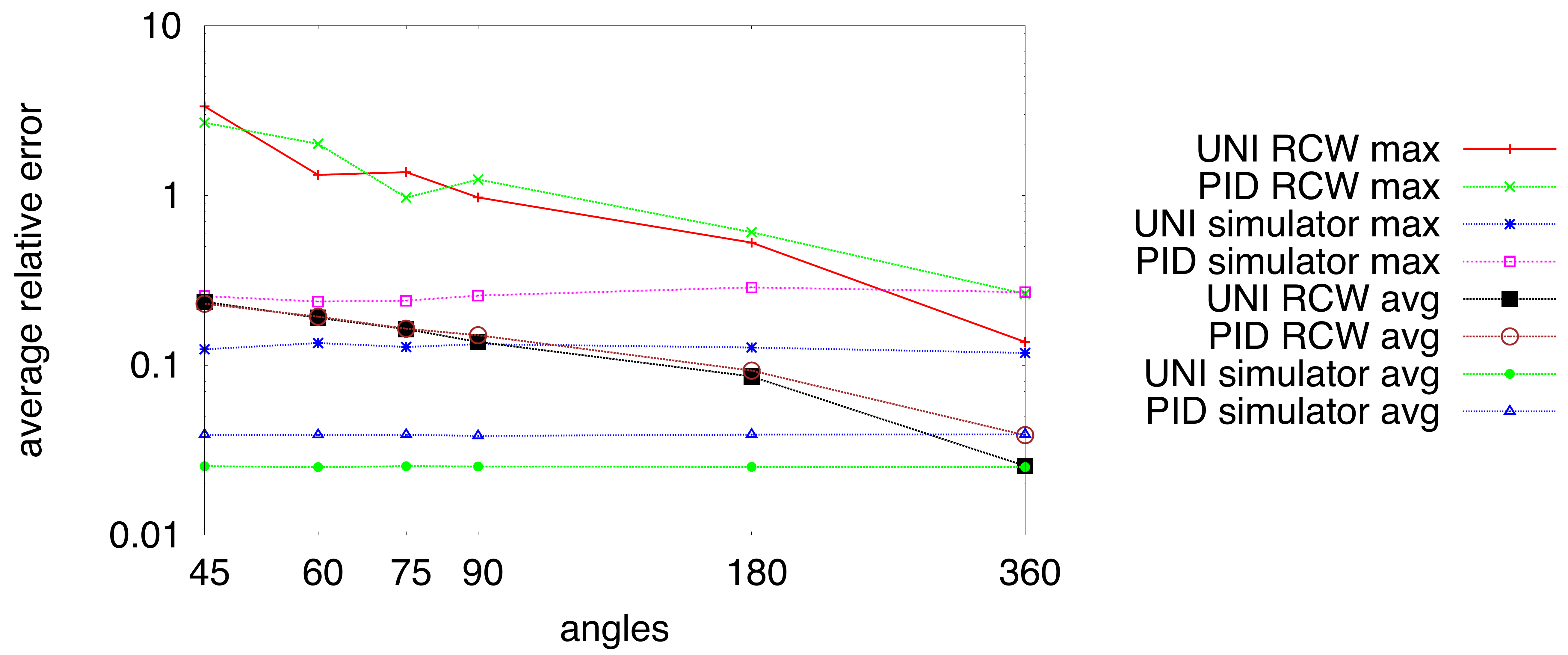}}
\vspace{-1.0em}
\caption{UNI and PID scenarios without uniform connectivity.}
\label{fig:E1-2-left}
\end{figure*}

\begin{figure}[t]
\begin{minipage}{0.70\textwidth}
%\lstset{basicstyle=\footnotesize}
\begin{lstlisting}
function $\mathit{AssignAttachmentPoints} (x,k)$
$ap$:= $\frac{\mathit{LCM}(n_k,n_{k+1})}{n_k}$
$C$:=$\mathbb{N}_{k+1}(x)$    /* neighbors of $x$ in ring $k+1$ */
$A_x$:= $\emptyset$   /* $A_x$ is a multiset */
loop
  choose $c$ from $C$
  $\mathit{send}$ ATTACH_MSG to $c$
  $\mathit{receive}$ RESPONSE_MSG from $c$ 
  if RESPONSE_MSG = OK then
    $ap$ := $ap - 1$
    add $c$ to $A_x$       /* $c$ can be in $A_x$ several times */
  else $C$ := $C \setminus \{c\}$
until $(ap = 0) \lor (C = \emptyset)$
if $(ap = 0)$ then return $A_x$
else return FAILURE
\end{lstlisting}
\end{minipage}
\hfill
\begin{minipage}{0.25\textwidth}
\begin{tabular}{|c|c|}
\hline
 Angle & \% success \\
\hline
\hline
 $15\,^{\circ}$ &0\%\\
 \hline
 $30\,^{\circ}$ & 0\%\\
  \hline
 $45\,^{\circ}$ & 3\%\\
  \hline
 $60\,^{\circ}$ & 82\%\\
  \hline
 $75\,^{\circ}$ & 99\%\\
  \hline
 $90\,^{\circ}$ & 100\%\\
  \hline
 $150\,^{\circ}$ & 100\%\\
  \hline
 $180\,^{\circ}$ & 100\%\\
  \hline
 $360\,^{\circ}$ & 100\%\\
\hline
\end{tabular}
\end{minipage}

\vspace{-1em}
\caption{Assignment Attachment Points (AAP) Function (left side). Success rate of the AAP algorithm as a function of the connectivity angle (right side).}
\label{AAP}
\label{tab:SuccessHalls}
\end{figure}

\begin{figure*}[t]
\scalebox{0.22}[0.22]{\includegraphics{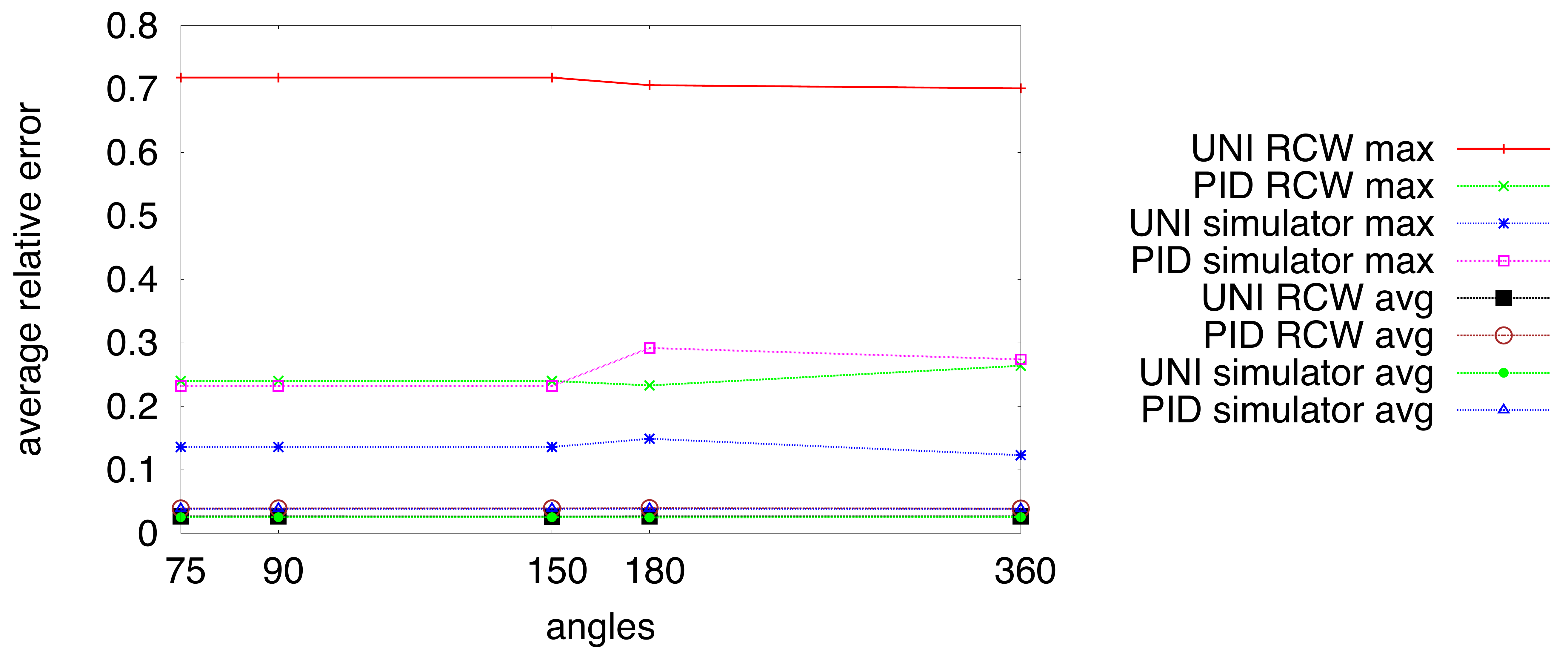}}
\vspace{-1.0em}
\caption{UNI and PID scenarios without uniform connectivity, using the AAP algorithm.}
\label{fig:E1-2-right}
\end{figure*}

Figure \ref{fig:E1-2-left} presents the results obtained in the UNI and PID scenarios. In both cases, we can see that the RCW algorithm performs much worse than the UNI and PID simulators. 
The simulation results show a biased behavior of RCW algorithm because the condition of Eq.~\ref{satisfy_cond} is not fulfilled in this experiment (i.e. a node has no neighbors, or there are two nodes in a ring $k$ that have different number of neighbors in rings $k-1$ or $k+1$). 

%This is mainly due to the fact that there are nodes without neighbors in the next ring except when large angles ($\ge180^{\circ}$) are used.

\vspace{-1em}
\paragraph{Assignment Attachment Points (AAP) Algorithm}

To eliminate the errors observed when there is no uniform connectivity, we propose a simple algorithm to transform the concentric rings network without uniform connectivity into an overlay network with uniform connectivity. 

To preserve the property that the visit probability is the same for all the nodes in a ring, nodes will use different probabilities for different neighbors. Instead of explicitly computing the probability for each neighbor, we will use the following process. Consider rings $k$ and $k+1$. Let $r=\mathit{LCM}(n_k,n_{k+1})$, where $\mathit{LCM}$ is the {\em least common multiple} function. 
We assign $\frac{r}{n_k}$ \emph{attachment points} to each node in ring $k$, and $\frac{r}{n_{k+1}}$ attachment points to each node in ring $k+1$. Now, the problem is to connect each
attachment point in ring $k$ to a different attachment point in ring $k+1$ (not necessarily in different nodes). If this can be done, we can use the algorithm of Figure \ref{f-drw}, but when a RCW is sent to the next ring, an attachment point (instead of a neighbor) is chosen uniformly. Since the number of attachments points is the same in all nodes of ring $k$ and in all nodes of ring $k+1$, the impact in the visit probability is that it is again the same for all nodes of a ring.

The connection between attachment points can be done with the simple algorithm presented in Figure \ref{AAP}, in which a node $x$ in ring $k$ contacts its neighbors to request available  attachment points. If a neighbor that is contacted has some free attachment point, it replies with a response message  $\mathit{RESPONSE\_MSG}$ with value \emph{OK}, accepting the connection. Otherwise it replies to $x$ notifying that all its attachment points have been connected. The node $x$ continues trying until its $\frac{r}{n_k}$ attachment points have been connected or none of its neighbors has available attachment points. If this latter situation arises, then the process failed. \textcolor{red}{The algorithm finishes in $O(\max_k \{n_k\})$ communication rounds. (Note that $r \le n_k \cdot n_{k+1}$ and $|C| \le n_{k+1}$).}
Combining these results with the analysis of Section~\ref{s:uniform}, we can conclude with the following theorem.

\begin{theorem}
Using attachment points instead of links and the distributed RCW-based algorithm of Figure \ref{f-drw}, it is possible to sample a concentric rings network without  uniform connectivity with any desired distance-based probability distribution $p_k$, provided that the algorithm of Figure \ref{AAP} completes (is successful) in all the nodes.
\end{theorem}

Figure \ref{fig:E1-2-right} shows the results when using the AAP algorithm. 
As we can see, the differences have disappeared. The conclusion is that, when nodes are placed uniformly at random and AAP is used to attach neighbors to each node, 
RCW performs as good as perfect UNI or PID simulators.

In general, the algorithm of Figure \ref{AAP} may not \textcolor{red}{be succesful}. 
%(However, it can be shown that it is always successful if there is uniform connectivity.)
It is shown in the table of Figure \ref{tab:SuccessHalls} (right side) the success rate of the algorithm for different connectivity angles.
It can be observed that the success rate is large
as long as the connectivity angles are not very small (at least $60^{\circ}$).
(For an angle of $60^{\circ}$ the expected
number of neighbors in the next ring for each node is less than 17.)
For small angles, like $15^{\circ}$ and $30^{\circ}$, the AAP algorithm is never successful.
For these cases, 
the algorithm for connected network presented in Section \ref{s:spt-rings} can be used.

%\input{su-simulations}

% Del RCW con arbol compartido
%\input{nuevademo}

\section{Conclusions}
\label{s:conclusions}

In this paper we propose distributed algorithms for node sampling in networks. All the proposed algorithms are based on a new class of random walks called centrifugal random walks. These algorithms guarantee that the sampling end after a number of hops upper bounded by the diameter of the network, and it samples with the exact probability distribution.
As future works we want to explore sampling in dynamic networks using random centrifugal walks.
\textcolor{red}{Additionally, we will investigate a more general algorithm that would also concern distributions that do not only depend on the distance from the source}.

%\bibliographystyle{plain}

%\bibliography{referencias}

%\newpage

%\begin{appendices}
%\appendix

%\input{appendix}

%\input{larger-distances}

%\end{appendices}

%\input{simulations}

%\input{protocol-jl}

\end{document}